\documentclass[final]{svjour3}
\usepackage{graphicx}
\usepackage{wrapfig}
\usepackage{rotating}
\usepackage{amsmath}
\usepackage{amssymb}
\usepackage{mathtools}
\usepackage{xfrac}
\usepackage[cal=cm]{mathalfa}
\usepackage[numbers,square]{natbib}
\usepackage{textgreek}
\usepackage{xcolor}

\makeatletter
\let\cl@chapter\undefined
\makeatletter
\usepackage[capitalise]{cleveref}

\usepackage{booktabs}
\usepackage{multirow}
\usepackage{textgreek}
\usepackage{siunitx}
\usepackage{xspace}
\usepackage{physics}

\makeatletter

\journalname{Journal of Low Temperature Physics}

\bibpunct{}{}{,}{s}{}{,}

\begin{document}
\newcommand{\hdblarrow}{H\makebox[0.9ex][l]{$\downdownarrows$}-}
\newcommand{\Ricochet}{\textsc{Ricochet}\xspace}
\newcommand{\cevns}{CE$\nu$NS\xspace}

\title{\Ricochet Progress and Status}

\authorrunning{C.~Augier et al.} 
\author{$^{1}$C.~Augier\and 
$^{1}$G.~Beaulieu\and 
$^{9}$V.~Belov\and 
$^{2}$L.~Berge\and 
$^{1}$J.~Billard\and 
$^{3}$G.~Bres\and 
$^{3}$J-.L.~Bret\and 
$^{2}$A.~Broniatowski\and 
$^{3}$M.~Calvo\and 
$^{1}$A.~Cazes\and 
$^{1}$D.~Chaize\and 
$^{2}$M.~Chapellier\and 
$^{7}$L.~Chaplinsky \and 
$^{4}$G.~Chemin \and 
$^{5}$R.~Chen \and 
$^{1}$J.~Colas \and 
$^{1}$M.~De Jesus \and 
$^{2}$P.~de Marcillac \and 
$^{2}$L.~Dumoulin \and 
$^{3}$O.~Exshaw \and 
$^{1}$S.~Ferriol \and 
$^{5}$E.~Figueroa-Feliciano \and 
$^{1}$J.-B.~Filippini \and 
$^{6}$J.~A.~Formaggio \and 
$^{8}$S.~Fuard \and 
$^{1}$J.~Gascon \and 
$^{2}$A.~Giuliani \and 
$^{3}$J.~Goupy \and 
$^{4}$C.~Goy \and 
$^{1}$C.~Guerin \and 
$^{10}$ C.~F.~Hirjibehedin \and
$^{6}$P.~Harrington \and 
$^{6}$S.~T.~Heine \and 
$^{7,*}$S.A.~Hertel \and
$^{4}$M.~Heusch \and
$^{4}$C.~Hoarau \and
$^{11}$Z.~Hong \and 
$^{1}$J.-C.~Ianigro \and 
$^{12}$Y.~Jin \and 
$^{6}$J.P.~Johnston \and 
$^{1}$A.~Juillard \and 
$^{9}$S.~Kazarcev \and 
$^{4}$J.~Lamblin \and 
$^{1}$H.~Lattaud \and 
$^{9}$A.~Lubashevskiy \and 
$^{6}$D.~W.~Mayer \and 
$^{2}$S.~Marnieros \and 
$^{3}$J.~Minet \and 
$^{1}$D.~Misiak \and 
$^{3}$A.~Monfardini \and 
$^{1}$F.~Mounier \and 
$^{2}$E.~Olivieri \and 
$^{2}$C.~Oriol \and 
$^{7}$P.K.~Patel \and 
$^{4}$E.~Perbet \and
$^{7}$H.D.~Pinckney \and 
$^{9}$D.~Ponomarev \and 
$^{2}$D.~Poda \and 
$^{4}$F.~Rarbi \and
$^{4}$J.-S.~Real \and 
$^{4}$J.-S.~Ricol \and
$^{2}$T.~Redon \and 
$^{8}$A.~Robert \and 
$^{9}$S.~Rozov \and 
$^{9}$I.~Rozova \and 
$^{1}$T.~Salagnac \and 
$^{1}$V.~Sanglard \and 
$^{5}$B.~Schmidt \and 
$^{9}$Ye.~Shevchik \and 
$^{6,1}$V.~Sibille \and 
$^{8}$T.~Soldner \and 
$^{6}$J.~Stachurska \and 
$^{4}$A.~Stutz \and 
$^{1}$L.~Vagneron \and 
$^{6}$W.~Van De Ponteseele \and 
$^{4}$F.~Vezzu \and
$^{10}$S.~Weber \and
$^{6}$L.~Winslow \and 
$^{9}$E.~Yakushev \and 
$^{9}$D.~Zinatulina }

\institute{$^{1}$Univ Lyon, Université Lyon 1, CNRS/IN2P3, IP2I-Lyon, F-69622, Villeurbanne, France \\
$^{2}$Université Paris-Saclay, CNRS/IN2P3, IJCLab, 91405 Orsay, France\\ 
$^{3}$Univ. Grenoble Alpes, CNRS, Grenoble INP, Institut Néel, Grenoble, France 38000\\
$^{4}$Univ. Grenoble Alpes, CNRS, Grenoble INP, LPSC-IN2P3, Grenoble, France 38000\\
$^{5}$Department of Physics, Northwestern University, IL, USA\\ 
$^{6}$Laboratory for Nuclear Science, Massachusetts Institute of Technology, Cambridge, MA, USA 02139\\ 
$^{7}$Department of Physics, University of Massachusetts at Amherst, Amherst, MA, USA 02139\\
$^{8}$Institut Laue Langevin, Grenoble, France 38042\\
$^{9}$Department of Nuclear Spectroscopy and Radiochemistry, Laboratory of Nuclear Problems, JINR, Dubna, Moscow Region, Russia 141980\\
$^{10}$MIT Lincoln Laboratory, Lexington, MA, USA\\
$^{11}$Department of Physics, University of Toronto, ON, Canada\\
$^{12}$C2N, CNRS, Univ. Paris-Saclay, 91120 Palaiseau, France\\
$^*$\email{shertel@umass.edu}\\
}

\maketitle

\begin{abstract}
    
We present an overview of recent progress towards the \Ricochet coherent elastic neutrino nucleus scattering (\cevns) experiment.  The ILL research reactor in Grenoble, France has been selected as the experiment site, after \textit{in situ} studies of vibration and particle backgrounds.  We present background rate estimates specific to that site, along with descriptions of the planned CryoCube and Q-Array detector payloads.


\end{abstract}

\section{Introduction}\label{sec1}

Coherent elastic neutrino nuclear scattering (\cevns) was predicted in 1974~\cite{PhysRevD.9.1389} and observed experimentally for the first time in 2017~\cite{doi:10.1126/science.aao0990}.  With first observation in hand, the next challenge is precision rate and spectral measurements, including searches for new physics.  While the initial \cevns searches took advantage of an increasing cross section at higher neutrino energy, new physics signatures are often aided by the use of comparatively low-energy neutrinos, where for example neutrino electromagnetic moments can appear as an increase in the \cevns rate~\cite{ricochetchooz}.  These searches for new physics in the neutrino sector, together with the practical goal of nuclear reactor monitoring, motivate the as-yet unachieved goal of \cevns observations using reactor neutrinos as the source.  A precision measurement of reactor neutrino \cevns spectra will require 1) a kg-scale target mass, 2) low background rates in a challenging environment, 3) thresholds of \SI{\approx 50}{\eV} or lower to capture a significant fraction of the \cevns spectrum, and lastly 4) the ability to reject electron-recoil backgrounds on an event-by-event basis at this low threshold.  This last point is one of the main distinctions of the \Ricochet approach.  The \Ricochet collaboration is currently engaged in an aggressive effort to meet all four of these requirements in a short timescale, through a combination of technologies.

\begin{figure}[htbp]
		\vspace{-0.1cm}
    \begin{center}
		\includegraphics[width=0.8\linewidth, keepaspectratio]{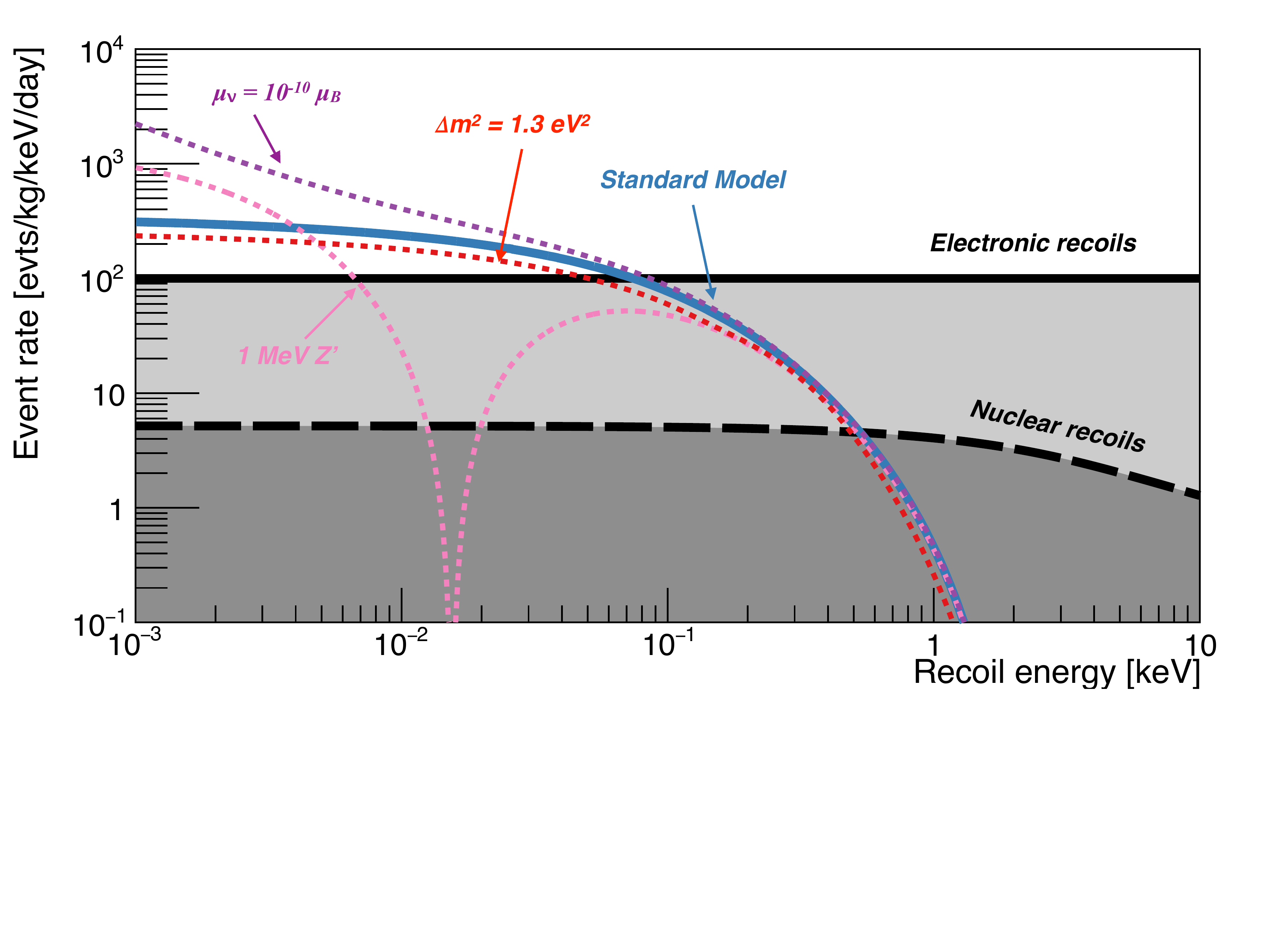}
	\end{center}
		\vspace{-2.2cm}
	\caption{Expected event rate and targeted background levels as a function of the recoil energy for the \Ricochet{} experiment deployed  8.8 meters from the ILL reactor core. The blue solid line is the standard model predicted \cevns event rate while the pink, purple and red dotted lines are respectively from adding a \SI{1}{\MeV} $Z_0$ boson ($g_{Z’} = 10^{-5}$), a neutrino magnetic moment of $\mu_\nu = 10^{-10}\mu_B$, and a sterile neutrino with $\Delta m^2 = \SI{1.3}{\eV^2}$ ($\sin^2(2\theta) = 0.5)$. The black solid and long-dashed lines represent the electronic and nuclear recoil background targeted levels respectively. Figure adapted from~\cite{Billard:2018jnl}.}
	\label{fig:RicochetCENNSNuPhys}
\end{figure}

\section{The ILL Site}\label{sec2}

\begin{figure}[htbp]
    \begin{center}
		\includegraphics[width=0.8\linewidth, keepaspectratio]{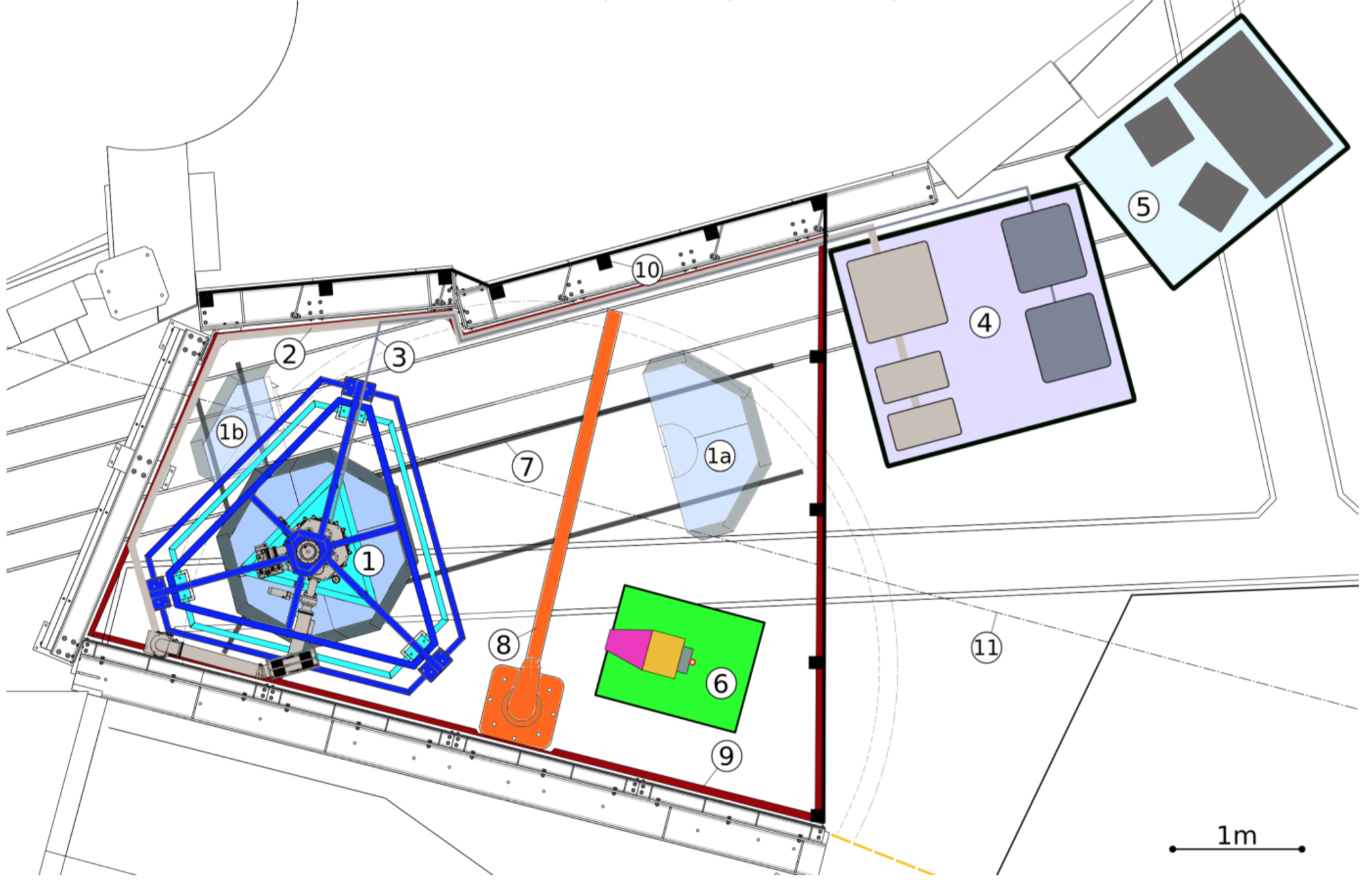}
	\end{center}
	\caption{Schematic of the \Ricochet{} installation in the H7 experimental area: 1) cryostat with shielding, electronics and double frame, 2) supply and primary vacuum lines, 3) data readout lines, 4) technical cabin for cryostat infrastructure and data acquisition servers, 5) control cabin, 6) pulsed neutron source (storage position), 7) rail system, 8) local 1-t crane, 9) retention walls, 10) light \Ricochet{} casemate, 11) limit of reactor transfer channel. The muon veto (not shown) is located above and around the cryostat and shielding.}
	\label{fig:RicochetILL}
\end{figure}

The future \Ricochet{} experiment will be deployed at the ILL-H7 site,  see \cref{fig:RicochetILL} for a preliminary schematic of the \Ricochet{} installation. The H7 site starts at about \SI{8}{\m} from the ILL reactor core that provides a nominal nuclear power of \SI{58.3}{\MW}, leading to a neutrino flux at the \Ricochet{} detectors \SI{8.8}{\m} from the reactor core of about 1.2$\times$10$^{12}$~cm$^{-2}$s$^{-1}$. The reactor is operated in cycles of typically 50 days’ duration with reactor-off periods sufficiently long to measure reactor-independent backgrounds, such as internal radioactivity or cosmogenic induced backgrounds, with high statistics. The available space is about \SI{3}{\m} wide, \SI{6}{\m} long and \SI{3.5}{\m} high. It is located below a water channel providing about \SI{15}{\m} water equivalent (m.w.e.) against cosmic radiations. It is not fed by a neutron beam and is well-shielded against irradiation from the reactor and neighboring instruments (IN20 and D19). The site is well-characterized in terms of backgrounds, and the operation of the STEREO neutrino experiment at this site has been successfully demonstrated~\cite{STEREO:2018blj}.

\section{Cryostat and Shielding}\label{sec3}

\begin{figure}[htbp]
    \begin{center}
		\includegraphics[width=0.88\linewidth, keepaspectratio]{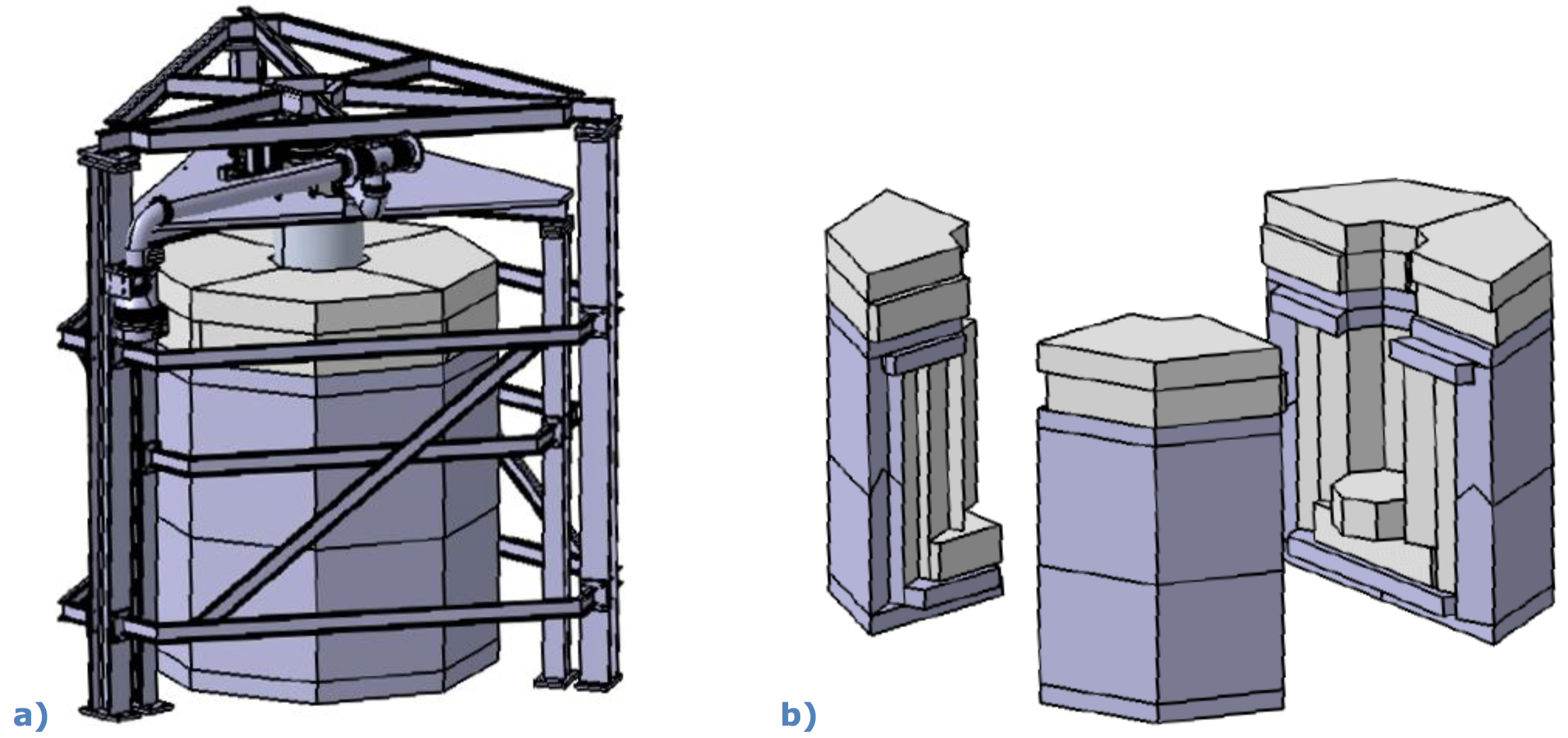}
	\end{center}
	\caption{\Ricochet{} shielding: a) Closed shielding inside the cryostat support frame. b) In order to retract the shielding, the horizontal bars of the support frame are removed. The shielding consists of three parts (2$\times$90$^\circ$ + 1$\times$180$^\circ$) that can be moved on a rail system (see \cref{fig:RicochetILL}).}
	\label{fig:RicochetShielding}
\end{figure}

A Hexa-Dry 200 Ultra-quiet cryostat from the CryoConcept Company, France, will be used. It is a dry cryogen-free cryostat with cold stages at \SI{50}{\K}, \SI{4}{\K}, \SI{1}{\K}, \SIlist{100;10}{\milli \K}. The \SI{10}{\milli \K} stage holds the detectors. The \SI{1}{\K} stage provides enough cooling power for the cold front-end electronics and also holds the cold shielding. Although this electronics is thermally anchored to the \SI{1}{\K} stage, it is located close to the detectors in order to minimise the stray capacitance from the cabling. The warm electronics is directly mounted on the \SI{300}{\K} flange. Since cryogenic detectors are very sensitive to vibrations, the ``Ultra Quiet Technology'' option was chosen. The cold head and its rotating valve are decoupled from the dilution unit. The cryostat is mounted on the inner and the mechanical parts (including pumps) on the outer layer of a double frame support structure. All vacuum connections between these elements are by edge-welded bellows only. A comparison has shown that this technology provides the smallest vibration level of the compared cryostats~\cite{Olivieri:2017lqz}. The cryostat support is exposed to vibrations of the floor of the H7 area and to acoustic noise in ILL. Vibration measurements have shown that excesses above \SI{\approx50}{\Hz} have to be suppressed by one-to-two orders of magnitude for optimal operation of the detectors. Simplified test measurements and preliminary estimates indicate that this can be achieved by insulating the inner frame from the floor passively with visco-elastic materials.

The background level at the ILL site is expected to be high. This is due to the proximity to the nuclear reactor core ($\approx$8~meters), the neighboring experiments emitting large amounts of gammas and neutrons (IN20 and D19), and the fact that, despite of the $\approx$15 m.w.e. artificial overburden provided by the water transfer channel of the reactor directly above the experiment, the site remains exposed to cosmic irradiation. As we expect to observe about 12.8 \cevns events/kg/day, a highly efficient background mitigation strategy is mandatory. As a matter of fact we are aiming for an electronic recoil background at the level of 100~events/day/keV/kg as those will be efficiently rejected thanks to our detectors' particle identification capabilities. However, as such discrimination doesn't hold for neutron induced nuclear recoil, the latter are expected to be our ultimate background. We are therefore aiming for a nuclear recoil background level around 5~events/day/kg to ensure a \cevns signal to noise ratio greater than one.  The \Ricochet{} shielding will be divided into two parts: a \SI{300}{\K} outer shielding illustrated in \cref{fig:RicochetShielding}, and a cold inner one (not shown). The outer passive shielding consists of \SI{20}{\cm} lead and \SI{35}{\cm} borated polyethylene. The cold inner shielding consists of \SI{13}{\cm} Pb/Cu and 8 layers of \SI{2.75}{\cm} polyethylene and 1~cm copper each.
Additionally, a third (external) \SI{35}{\cm} thick layer of polyethylene on top of the outer lead layer and \SI{8}{\mm} thick polyethylene layers mounted on each thermal screens will be used to reduce neutron spallation in the lead shielding and further improve the shielding tightness, respectively. As shown on \cref{fig:RicochetShielding}, the outer shielding will be made of three parts which will be installed on rails to allow accessing the cryostat. Lastly, muon induced gamma and neutron backgrounds will be further reduced thanks to a surrounding muon-veto allowing rejecting events in temporal coincidence with muons going through the experimental setup. Note that the cryostat will also be hosting muon veto panels anchored at its \SI{50}{\K} stage to cover for the hole from the cryostat imprint ensuring an almost full coverage of the experimental setup.

\section{Expected backgrounds}\label{sec4}

\newcommand{\TitleTabER}{\multirow{3}{*}{\begin{tabular}[c]{@{}c@{}}{\bf Electronic recoils} \\ {[}50\,eV, 1\,keV{]} \\ (evts/day/kg)\end{tabular}}}
\newcommand{\TitleTabNR}{\multirow{3}{*}{\begin{tabular}[c]{@{}c@{}}{\bf Neutron recoils}    \\ {[}50\,eV, 1\,keV{]} \\ (evts/day/kg)\end{tabular}}}

\newcommand{\CosmoNoShieldER}{$260 \pm 5$}                   \newcommand{\ReactoNoShieldER}{$4365 \pm 301$}               \newcommand{\SumNoShieldER}{$4625 \pm 301$}                 
\newcommand{\CosmoShieldER}{$183 \pm 6$}                     \newcommand{\ReactoShieldER}{$18\pm 2$}                     \newcommand{\SumShieldER}{$201 \pm 6$}                      
\newcommand{\CosmoShieldVetoER}{$1.6\pm0.6$}                                                                            \newcommand{\SumShieldVetoER}{{\bf 20$\pm$2}}                   

\newcommand{\CosmoNoShieldNR}{$1554 \pm 12$}                 \newcommand{\ReactoNoShieldNR}{$53853 \pm 544$}              \newcommand{\SumNoShieldNR}{$55407 \pm 545$}                
\newcommand{\CosmoShieldNR}{$42\pm 3$}                      \newcommand{\ReactoShieldNR}{$2.4\pm0.3$}                  \newcommand{\SumShieldNR}{$44\pm3$}                       
\newcommand{\CosmoShieldVetoNR}{$7\pm2$}                                                                               \newcommand{\SumShieldVetoNR}{{\bf 9$\pm$1}}

\begin{table}[!t]
\def\arraystretch{1.5}
\setlength{\tabcolsep}{1em}
\begin{tabular}{@{}ccccc@{}}
\multicolumn{2}{c}{}                            & Cosmogenic         & Reactogenic                       & Total         \\
\midrule[0.5pt]
 \TitleTabER & No Shielding (I)                    & \CosmoNoShieldER   & \ReactoNoShieldER                 & \SumNoShieldER                            \\
             & Passive Shielding (II)                & \CosmoShieldER     & \multirow{2}{*}{\ReactoShieldER}  & \SumShieldER                               \\
             & Passive + $\mu$-veto (III) & \CosmoShieldVetoER &                                   & \SumShieldVetoER            \\
\midrule[0.5pt]
 \TitleTabNR & No Shielding  (I)                   & \CosmoNoShieldNR   & \ReactoNoShieldNR                 & \SumNoShieldNR                            \\
             & Passive Shielding (II)               & \CosmoShieldNR     & \multirow{2}{*}{\ReactoShieldNR}  & \SumShieldNR                                \\
             & Passive + $\mu$-veto (III) & \CosmoShieldVetoNR &                                   & \SumShieldVetoNR             \\
\midrule[0.5pt]
{{\bf \cevns}} & & & & {\bf 12.8} \\
\bottomrule[0.5pt]
\end{tabular}
\caption{Simulated background rates inside the cryogenic detector array, with the preliminary shielding design, when only one bolometer has triggered. As the muon-veto is still being characterized and optimized, we  assume here perfect geometrical and detection efficiencies.}
\label{tab:SimuRate}
\end{table}

The \Ricochet background model includes both cosmogenic and radiogenic backgrounds which are respectively generated and propagated to our \Ricochet Geant4-based simulation using respectively the CRY cosmic ray generator, and MCNP simulations combined with onsite gamma and neutron spectrometer measurements~\cite{valerian}. Note that we do not consider here the radiogenic background component as it is currently under investigation, combining ongoing material screening measurements and Geant4 simulations, and is expected to be sub-dominant. The resulting rates for both electronic and nuclear recoil backgrounds from cosmogenic and reactogenic origins, and under different shielding configurations, are presented in \cref{tab:SimuRate}. We found that while the total electronic background is well below its targeted level, the neutron one is about twice its targeted level (assuming a 100\% efficient muon veto), implying an expected S/B$\approx$1. We see that our dominating source of nuclear recoil background is from the muon induced neutrons, and by comparing the results from configurations (II) and (III) we can conclude that a full muon-veto coverage is essential if 1) the spurious gamma induced trigger rate is low enough during reactor's operation, and 2) the induced dead time is reasonable. With an expected \SI{350}{\Hz} muon-veto trigger rate, and a demonstrated $\approx$100~$\mu$s timing resolution for the Ge cryogenic detectors thanks to the fast ionization signal, we anticipate a dead-time between 20-30\% while dividing the muon-induced gamma and neutron backgrounds by about 160 and 6, respectively~\cite{hdr_julien}. Interestingly we see that the expected \cevns signal is expected to be 5 times larger than the particularly damaging correlated reactogenic NR background of 2.4 events/day/kg. This is explained by the fact that despite of their large flux, about 10 times larger than the cosmogenic one~\cite{valerian}, due to their relatively low maximum energy of about \SI{10}{\MeV} reactogenic neutrons are  highly efficiently stopped within our passive shielding layers.

\section{CryoCube}\label{sec5}

\subsection{Overview}

\begin{figure}[htbp]
    \begin{center}
		\includegraphics[width=0.358\linewidth, keepaspectratio]{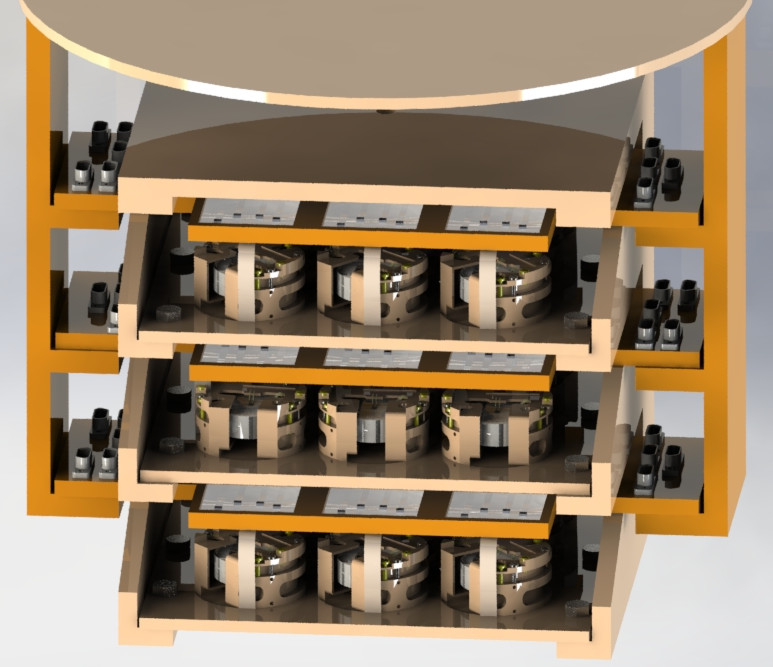}
		\includegraphics[width=0.30\linewidth, keepaspectratio]{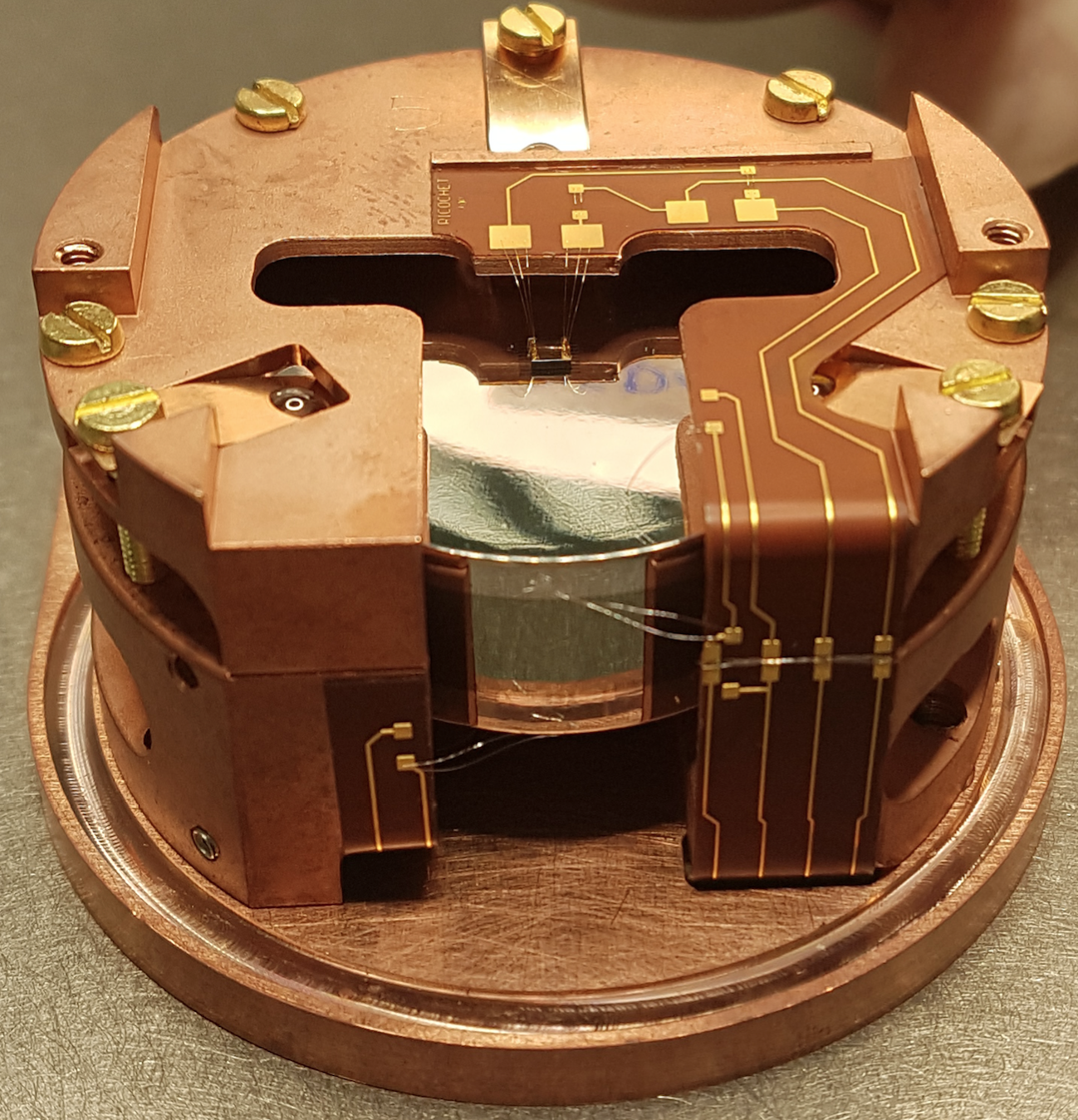}
		\includegraphics[width=0.29\linewidth, keepaspectratio]{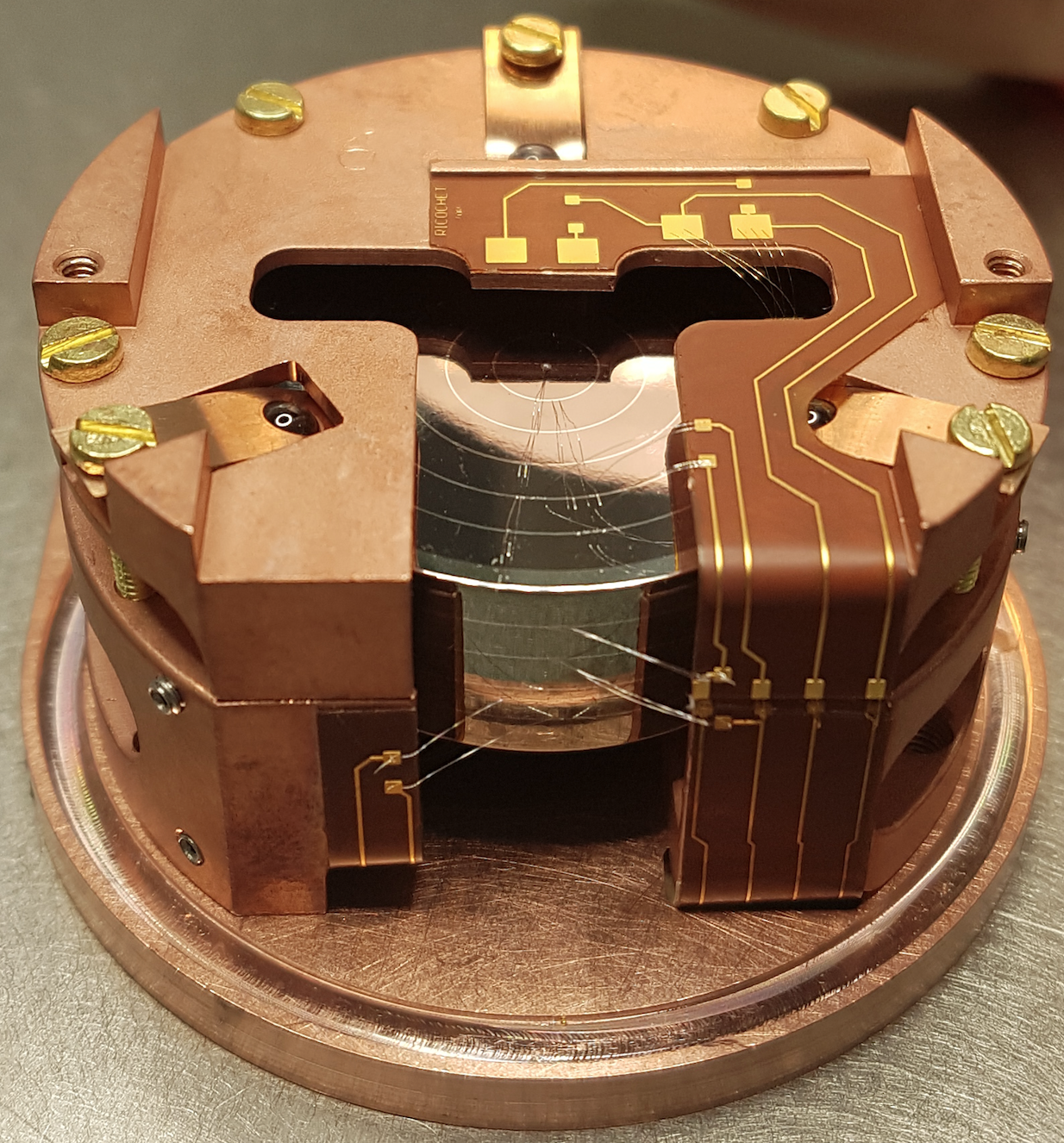}
	\end{center}
	\caption{{\bf Left:} Draft model of the CryoCube design. All the parts in dark orange correspond to 1\,K stage holding the HEMT-based electronics. The light part supports the 27 Ge detectors  regulated at about 20\,mK. Pictures of a planar ({\textit middle}) and an FID ({\textit right}) detector prototype.}
	\label{fig:CryoCube}
\end{figure}

The CryoCube will consist of an array of 27 ($3\times3\times3$) high purity germanium crystal detectors, encapsulated in a radio-pure infrared-tight copper box suspended below the inner shielding, see \cref{fig:CryoCube} (left panel). Each detector mass is about 38\,g to reach a total target mass around one kilogram. A $\order{\SI{10}{\eV}}$ energy threshold is desired as the discovery potential scales exponentially with lowering the energy threshold. Considering a 50\,eV energy threshold, about 12.8\,evts/day of \cevns interactions is expected in the CryoCube detector array.
To reach such threshold, the CryoCube detectors will be equipped with germanium neutron transmutation doped sensors (NTD). To achieve particle identification, the detectors will have a double heat and ionization readout. Ionization measurement is realized thanks to aluminum electrodes allowing to apply an electric field and collect signals from the ionization electron-hole pairs drifting across the crystal. With an anticipated particle identification threshold of about 100\,eV, thanks to the combination of a 10\,eV and 20\,eVee (electron-equivalent) heat and ionization baseline resolutions (RMS), the CryoCube detector array should lead to a \cevns detection significance after one ILL reactor cycle (50-days) between 4.3--17.3~$\sigma$, depending on the final background level achieved~\cite{LTD_Cryocube}.

\subsection{Heat channel optimization}

A comprehensive electro-thermal model has been developed to optimize the heat energy resolution of the CryoCube germanium detectors which are equipped with a single NTD heat sensor~\cite{These_Dimitri}. The individual detector mass has been first validated with 33.4\,g germanium prototypes with an average heat channel baseline resolution of 22\,eV (RMS) on five detectors. The best resolution achieved, 17\,eV (RMS), was obtained with the modulated JFET-based EDELWEISS electronics at surface level~\cite{EDELWEISS:2019vjv}. Thanks to their 30\,g-scale mass and 20\,eV (RMS) resolution, CryoCube detector prototypes have proven to be expecting the highest \cevns event rate per crystal ($\approx$0.3\,events/day) with respect to the current cryogenic detector state-of-the-art from surface operation. Lastly, to achieve the CryoCube specifications, {\it i.e.} 10\,eV (RMS) heat baseline resolution and a resulting \cevns rate per crystal of $\approx$0.6\,events/day, we are developing dedicated low-noise HEMT-based preamplifiers~\cite{Juillard2020, JLTP_HEMT}.

\subsection{Ionization channel optimization}

Two electrode designs are being considered for the CryoCube: planar detector (PL) shown in \cref{fig:CryoCube} (middle panel), with one electrode on the top and another one on the bottom of the crystal, and fully interdigitated detector (FID) (right panel), with ring electrodes covering the entire crystal surfaces. While the PL design offers a larger fiducial volume and improved charge collection thanks to its electric field uniformity, it also lacks from the surface event rejection inherent to the FID design~\cite{JINST_EDW_2017} . Both designs have therefore their own pros and cons, and a definite choice as to the which is best for the CryoCube array will be done following first in-situ background measurements. For the time being, both designs are being further studied, optimized and tested. While their charge collection capabilities differ, both designs need to have low capaticance electrodes ({\it i.e.} less than \SI{20}{\pico \F}) to ensure a 20~eVee (RMS) ionization baseline resolution with our upcoming HEMT-based preamplifiers. Indeed, while for the heat channel switching from a JFET- to a HEMT-based electronics only improve the resolution by a factor of two, it should improve our ionization resolution by a factor of 8. This is due to the combination of both a much lower HEMT intrinsic current noise, and a ten fold reduced input capacitance summing the detector, cabling, and the HEMT gate (\SI{4.6}{\pico \F}) contributions~\cite{Juillard2020, JLTP_HEMT}. In addition to its ultra-low noise level our  HEMT-based charge preamplifier has been designed to be linear up to tens of MeV with a percentage-level gain stability over the 5 order of magnitude dynamic range. Lastly, to fulfill the timing constraints from the Ricochet muon veto and its expected \SI{\approx 350}{\Hz} muon induced triggering rate, the preamplifier has been designed to achieve a \SI{40}{\kHz} bandwidth hence allowing a combined heat/ionization timing resolution of 100$\mu$s (RMS) at \SI{100}{\eV}~\cite{hdr_julien,LTD_MPS}. A preamplifier prototype is currently being commissioned and should be first tested with a CryoCube detector prototype by Early-2022.

\section{Q-Array}\label{sec6}

\subsection{Overview}

Q-Array -- the complimentary detector array within Ricochet -- will consist of 9 cubes of superconducting zinc as its target. Using superconductors as the primary detector is a novel technology which is expected to provide a detection threshold theoretically down to the Cooper pair binding energy, and excellent background discrimination.  The expected discrimination mechanism begins with the different efficiency of quasiparticle (QP) production (breaking Cooper pairs) by electron recoils (higher QP production) vs. nuclear recoils (lower QP production).  The initial athermal phonon production is followed by a slower phonon production as QPs relax to the ground state.  The relative ratio between initial phonons and QP-induced phonons thereby gives rise to a usefully discriminating pulse shape.  As QP lifetime increases exponentially below their critical temperature (\SI{850}{\milli\K} for Zn) due to the suppressed coupling to phonons, we expect significant differences in the thermalization time constants between electronic and nuclear recoils. Transition edge sensors (TES) will be used for the readout of the athermal phonon signals from these superconducting bolometers. Initial prototype TES chips with a transition temperature of \SI{80}{\milli\K} have been developed by Argonne National Laboratory for this use. This technology is currently being pushed toward \SIrange{15}{20}{\milli\K}, which would result in a dramatic decrease in sensor threshold.  A prototype zinc detector has been fabricated at RMD, Inc.  Two gold pads (Zn-ZnO-Au and Zn-Au) provide a thermal link between the zinc crystal and the phonon/quasi-particle sensor.

\begin{figure}[htbp]
    \begin{center}
		\includegraphics[height=2.25in, keepaspectratio]{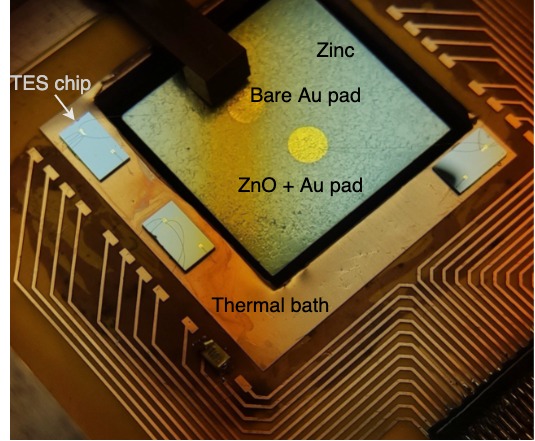}
		\includegraphics[height=2.25in, keepaspectratio]{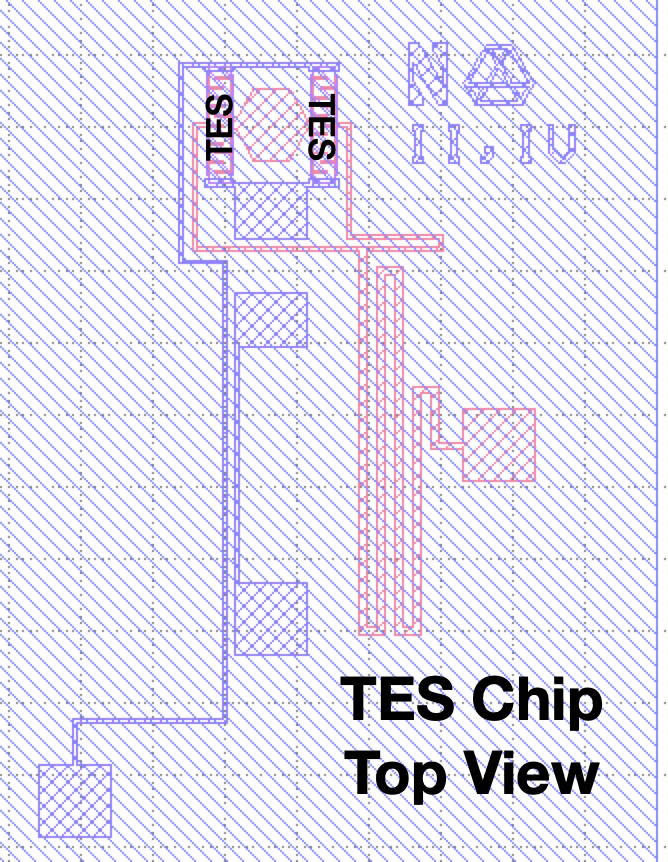}
	\end{center}
	\caption{{\bf Left:} Photo of a Zn superconducting target with Au pads, connected to three separated TES sensor chips using Au wire bonds. Additional gold wirebonds are used to connect the chips with the thermal bath and aluminium wirebonds to bias the TES. Individual wirebonds are too small to see.
	{\bf Right:} One design of TES thermometer chip. In this design two TES's share a single gold pad thermally connected to the absorber.  Purple indicates Al (readout leads), red indicates gold (thermal path into and out of the TES.}
	\label{fig:Qarray}
\end{figure}

\subsection{Heat channel optimization}

The Q-array design presented in a companion paper\cite{RanLTD} has been optimized using a block thermal model\cite{Fig:2006,tali2015}.  Heat flows from a large-area gold pad on the target through a gold wirebond to the separated TES sensor chip.  Then heat flows out of the TES to the thermal bath with a thermal conductivity set by an on-chip gold meander (red in \cref{fig:Qarray}, right).  The timescales of heat flow into and out of the TES are tuned (using a thermal model) to optimize the sensor's threshold and potential for pulse shape information.  This thermal modeling predicts a baseline energy resolution of \SI{15}{\eV} on a \SI{32}{\g} Ge target mass, assuming a \SI{40}{\milli\K} TES. Lowering the transition temperature to \SI{20}{\milli\K} will improve the energy resolution to achieve our \SI{50}{\eV} threshold goal. 

\subsection{Q-Array Readout}

The Q-array assembly will employ a separate readout scheme from the CryoCube array, whereby TES channels are read using a high frequency microwave multiplexer, akin to that used by the SLEDGEHAMMER X-ray calorimeter~\cite{sledgehammer}.  Aluminum-based RF-SQUID resonators serve as a first stage amplification for the low current signals from the TES channels, with each resonator tuned to a specific microwave frequency in a \SIrange{4}{7}{\GHz} window.  A prototype 6-channel Al-resonator chip has been fabricated at Lincoln Laboratory.  Tests are being conducted to see whether a broadband traveling wave parametric amplifier (TWPA) can be used as a second-stage amplification~\cite{TWPA}.

\section{Conclusion}\label{sec7}

The \Ricochet collaboration is engaged in several efforts proceeding in parallel.  As described in this manuscript, these efforts include the further improvement of shielding and veto geometry, the further optimization of the NTD- and HEMT-based CryoCube design to achieve a a low threshold with discrimination, and the testing of the complementary TES-based Q-Array design.  The shared goal of these efforts is to deploy at ILL a payload capable of statistically-significant \cevns detection, with first \cevns exposure starting in 2023.

\section*{Acknowledgments}

This project has received funding from the European Research Council (ERC) under the European Union’s Horizon 2020 research and innovation program under Grant Agreement ERC-StG-CENNS 803079, the French National Research Agency (ANR) within the project ANR-20-CE31-0006, the LabEx Lyon Institute of Origins (ANR-10-LABX-0066) of the Université de Lyon.  A portion of the work carried out at MIT was supported by DOE QuantISED award DE-SC0020181 and the Heising-Simons Foundation. A portion of this material is based upon work supported by the Under Secretary of Defense for Research and Engineering under Air Force Contract No. FA8702-15-D-0001. Any opinions, findings, conclusions or recommendations expressed in this material are
those of the author(s) and do not necessarily reflect the views of the Under Secretary of Defense for Research and Engineering. Authors are grateful for the technical and administrative support of the ILL.  Data sharing not applicable to this article as no datasets were generated or analysed during the current study.

\end{document}